\documentclass[namedreferences]{solarphysics}
\usepackage[optionalrh,solaenum,natbib]{spr-sola-addons} 
\usepackage{epsfig}                     
\usepackage{graphicx}                    
\usepackage{amssymb}                    
\usepackage{color}                       
\usepackage{url}                         
\usepackage[toc,page]{appendix}

\newcommand{\arcsec}{$''$}

\begin{document}

\begin{article}

\begin{opening}

\title{Off-limb (spicule) DEM distribution from SoHO/SUMER observations}

%
\author{K.~\surname{Vanninathan}$^{1,2}$\sep
        M. S.~\surname{Madjarska}$^{1}$\sep
        E.~\surname{Scullion}$^{3}$\sep      
        J. G.~\surname{Doyle}$^{1}$\sep
       }

%
\runningauthor{K. Vanninathan et al.}
\runningtitle{Off-limb (spicule) DEM distribution from SoHO/SUMER observations}

%
  \institute{$^{1}$ Armagh Observatory, College Hill, Armagh BT61 9DG, N. Ireland, UK\\
                     e-mail: \url{kva@arm.ac.uk} \\
                $^{2}$ School of Mathematics and Physics, Queen�s University Belfast, Belfast BT7 1NN, N. Ireland\\
                $^{3}$ Institute of Theoretical Astrophysics, University of Oslo, Norway\\
             }

\begin{abstract}
In the present work we derive a Differential Emission Measure (DEM) distribution from a region dominated by spicules. 
We use spectral data from the \textit{Solar Ultraviolet Measurements of Emitted Radiation} (SUMER)
spectrometer on-board the \textit{Solar Heliospheric Observatory} (SoHO) covering the entire SUMER wavelength range taken off-limb
in the Northern polar coronal hole to construct this DEM distribution using the CHIANTI atomic database. This distribution is
then used to study the thermal properties of the emission contributing to the 171~\AA\ channel in the \textit{Atmospheric Imaging Assembly} (AIA)
on-board the \textit{Solar Dynamics Observatory} (SDO).  From our off-limb DEM we found that
the radiance in the AIA 171~\AA\ channel is dominated by emission from the Fe~{\sc ix}~171.07~\AA\ line and has sparingly little
contribution from other lines. The product of the Fe~{\sc ix}~171.07~\AA\ line contribution function with the off-limb DEM was found to have a
maximum at log$T_{max}~(K)~=~5.8$ indicating that during spicule observations the emission in this line comes from plasma at
transition region temperatures rather than coronal. For comparison, the same product with a quiet Sun and prominence DEM
were found to have a maximum at
log$T_{max}~(K)~=~5.9$ and  log$T_{max}~(K)~=~5.7$, respectively. We point out that the interpretation of data
obtained from the AIA 171~\AA\  filter should be done with foreknowledge of the thermal nature of the observed phenomenon.
For example, with an off-limb DEM we find that only 3.6\% of the plasma is above a million degrees,
whereas using a quiet Sun DEM, this contribution rises to 15\%.
\end{abstract}

%
\end{opening}
%
 \section{Introduction}
 For a long time spicules \citep{Beckers1968} have been examined as structures that intermittently 
couple the chromosphere and the corona through a continuous ejection of mass flux and thereby 
causing heating. However, the coronal counterparts of these jets were never found \citep{Withbroe1983, 
Mariska1992}. Thus the idea of their direct contribution to coronal heating was shelved
\citep{Withbroe1983} until recently. \citet{DePontieu2007} revived this idea when they used off-limb 
coronal hole data to show that high velocity spicules (so called type II spicules) existing in the solar chromosphere 
exhibit only upward motions. \cite{RouppeVanderVoort2009} correlated the off-limb type~II spicules 
to on-disk Rapid Blueshifted Excursions (RBEs) and established that these features occur
ubiquitously on the Sun. Their disappearance from the Ca~{\sc ii}~H passband images taken with the \textit{Solar Optical Telescope} (SOT)
on-board the \textit{Hinode} spacecraft has been interpreted as heating which causes the singly ionised calcium to become
at least  doubly ionised. This subject received further excitement when \citet{DePontieu2011} correlated the SOT Ca~{\sc ii}~H spicules
with their coronal equivalent as seen in 171~\AA\ and 211~\AA\ bandpass images of the \textit{Atmospheric Imaging Assembly} (AIA) on-board
\textit{Solar Dynamics Observatory} (SDO). Recently, \citet{Madjarska2011} analysed three large spicules seen in SOT Ca~{\sc ii}~H
images and concluded that ``these spicules although very large and dynamic, are not present in spectral lines formed at
temperatures above 300\,000~K'' observed with the \textit{Extreme-ultraviolet Imaging Spectrometer} (EIS) on-board \textit{Hinode}. In addition,
their preliminary analysis of solar prominences, which have very similar plasma parameters as spicules with respect to temperatures and
densities, showed that  prominences  are seen in the AIA 171~\AA\ images while EIS  observations for one of these prominences
revealed an emission not higher than 400\,000~K  (Fe~{\sc viii}, log$T~(K)~\sim~5.6$). The authors suggested, therefore,
that the recent observations of spicules by \cite{DePontieu2011} in AIA/SDO 171~\AA\ and 211~\AA\  channels may come
from the existence of transition region emission in these passbands.

The AIA \citep{Lemen2012} on-board SDO \citep{Pesnell2012} has revolutionised our view of the Sun 
with its unprecedented cadence, high spatial resolution and large field-of-view. AIA has 10 passbands 
which cover a wide range of temperatures thus enabling the study of the solar atmosphere from the 
chromosphere to the corona. Unfortunately, due to the relatively large spectral widths of the AIA 
passbands, the resultant images are not spectrally pure. Although the emission from these filters 
are dominated by the intended primary ion, initial analysis show that the contribution from 
several adjacent spectral lines cannot be ruled out \citep{ODwyer2010}. For their analysis they used typical Differential Emission
Measure (DEM) distributions for a coronal hole, quiet Sun, active region and flaring plasma. By forward modeling the emission
in the AIA filters, \cite{Martinez-Sykora2011} concluded that the 131\AA, 171~\AA\ and 304~\AA\ passbands
have a negligible contribution from the non-dominant ions. The authors  used  3D MHD numerical simulations of the solar atmosphere
for regions representing ``coronal hole, and quiet Sun with hot emerging regions and with hot corona" to calculate the intensity (using CHIANTI)
of  spectral lines within the AIA passbands, and investigate the importance of non-dominant lines for the various solar coronal conditions.
Important issues related to the AIA passbands have been highlighted recently such as the presence of unidentified transition region and coronal
lines in the 211~\AA\ channel \citep{DelZanna2011}. From their study of active region loop footpoints they also suggest that since the
171~\AA\ channel has a significant contribution from cool plasma (log$T~(K)~<~5.7$),  observations in this channel should be treated
with extreme caution. 

Similar studies have been conducted for some of the earlier observations by \citet{Brooks2006} where they investigated the response functions
for the 171~\AA\ and 195~\AA\ channels of \textit{Extreme-ultraviolet Imaging Telescope} (EIT) on-board
the \textit{Solar and Heliospheric Observatory} (SoHO) and the \textit{Transition Region and Coronal Explorer} (TRACE)
using coordinated \textit{Coronal Diagnostic Spectrometer} (CDS) observations of the quiet Sun. The TRACE response functions were also studied by
\citet{DelZanna2003} and \citet{Cirtain2005} for active regions and flares, respectively. 

The overview given above shows that until now there has been no attempt made to understand the thermal response of the filters on
different telescopes to spicule observations. The time is now ripe to study this problem as the current high spatial and time resolution
observations make it possible to look at features like spicules, bright points, prominences \textit{etc.} in great detail. In the present paper we
aim to obtain for the first time a DEM distribution to study the thermal properties of emission registered through the AIA 171~\AA\
passband during off-limb activity above a coronal hole, dominated by spicules. The results obtained here could be applied to
observations of any phenomenon with emission at chromospheric and/or transition region temperatures with the 171~\AA\ channel of
AIA/SDO. The data used here are described in Section 2. The DEM analysis is given in Section 3. In Section 4 we describe and
discuss our results, and in Section 5 we state our conclusions.

\section{Observations}
\begin{figure}[ht]
\centering
\includegraphics[width=10cm]{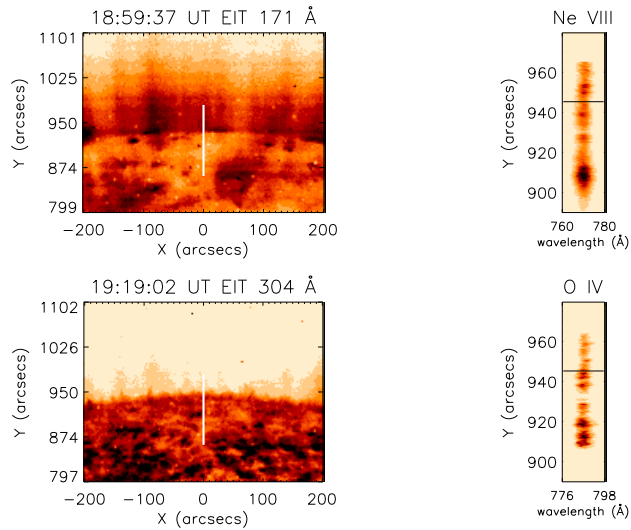}
\caption{\textbf{Top left}: EIT Fe~{\sc ix/x}~171~\AA\ context image with the SUMER slit over-plotted. 
\textbf{Bottom left}: EIT 304~\AA\ image of the same region showing spicular emission. The 
plotted solid line represents the position of the SUMER slit. \textbf{Top right}: Ne~{\sc viii}~770.42~\AA\ 
emission along the SUMER slit taken at 19:02:36 UT. \textbf{Bottom right}:  O~{\sc iv}~787.72~\AA\ along the SUMER 
slit taken at 19:02:36 UT. The horizontal line in both images denotes the limb position.}
\label{eit_sumer}
\end{figure}

The \textit{Solar Ultraviolet Measurements of Emitted Radiation} \citep[SUMER;][]{Wilhelm1995} instrument on-board SoHO \citep{Domingo1995}
is a telescope and a spectrometer which uses two detectors to cover a large spectral range from 465~\AA\ to 1610~\AA. While detector A is
sensitive to wavelengths 780~\AA\ to 1610~\AA, detector B works from 660~\AA\ to 1500~\AA. From the SUMER archive we selected suitable
data which will enable us to study transient events like spicules. The data were taken on 13~June~1998 in and above a coronal hole at the
North Pole. Coronal hole off-limb data are the most suitable for this study as the dominant features observed there are spicules.
\cite{RouppeVanderVoort2009} calculated that there are between 1.5 and 3 Ca~{\sc ii}~H type~II spicules per linear arcsec along the
limb in a coronal hole. Hence, our off-limb region of study is intermixed with both type~I and type~II spicules. In Figure~\ref{eit_sumer} 
we show context images in Fe~{\sc ix}~171~\AA\ (top, left) and He~{\sc ii}~304~\AA\  (bottom, left) taken with the EIT with the 
position of the SUMER slit over-plotted. We also show in the Appendix, Figure~\ref{aia_eit}, a comparison between EIT and AIA 304~\AA\ images in order 
to demonstrate that we are able to determine with EIT images various phenomena in coronal holes off-limb data, e.g. spicules and bright points, despite the instrumental limitations.

In Figure~\ref{eit_sumer}, we also show examples of the SUMER spectral data (top/bottom, right) taken co-temporally with the EIT images.
We used these
images to verify that the emission under the SUMER slit comes only from spicular material and no other features like, for instance, coronal
bright points, are present. The spectra were recorded on detector B with a slit of 0.3\arcsec~$\times$~120\arcsec\ and 300~s exposure time.
The SUMER data reduction was done with the standard software. The spectral  lines were identified with the help of the SUMER spectral atlas
\citep{Curdt2001}. A list of all the lines used, their corresponding formation temperatures and radiances are given in Table~\ref{T1}. We
selected all lines with formation temperatures between log$T~(K)~=~4.1$ and log$T~(K)~=~5.8$ which have a sufficient signal-to-noise ratio (above 10).
The Ne~{\sc viii} line was chosen as the upper limit since it is the best SUMER line which has high formation temperature (log$T~(K)~=~5.8$)
close to the formation temperature of Fe~{\sc ix}. All other SUMER lines with higher formation temperatures have in general very poor
signal-to-noise ratio. Furthermore, from the analysis of \cite{Madjarska2011} we know that spicules do not have any signal at temperatures higher than
0.3~MK which makes the use of only SUMER data for this study a reasonable approach. The alignment of the SUMER slit with the EIT images
was done by first selecting spectral lines in the SUMER data which have similar formation temperatures as the EIT channels (Fe {\sc ix} and
He {\sc ii}) and then by comparing them.

\begin{table}
\caption{Details of the SUMER spectral lines used for the construction of the DEM}
\label{T1}
\begin{tabular}{c c c c}
\hline\hline
Wavelength (\AA) & Ion & log~$T_{max}~(K)$ & Radiance $\times 10^{3} $\\
& & & $(ergs~cm^{-2} s^{-1} sr^{-1}$)\\
\hline
      765.150  &  	N~{\sc iv}       &  5.1   &  10.30\\
      770.420  &  Ne~{\sc viii}   &  5.8   &  6.32\\
      780.300  &	Ne~{\sc viii}   &  5.8   &  3.58\\
      786.470  &  	S~{\sc v}         &  5.2   &  4.82\\
      787.720  & 	O~{\sc iv} 	      &  5.2   &  8.89\\
      977.030  &  	C~{\sc iii}        &  4.8   &  179.98\\
      1031.93  &	O~{\sc vi}	      &  5.5   &  54.95\\
      1238.82  &  	N~{\sc v}         &  5.3   &  20.88\\
      1242.80  &	N~{\sc v}         &  5.3   &  10.68\\
      1253.80  &  	S~{\sc ii}         &  4.2   &  2.52\\
      1298.96  & 	Si~{\sc iii}       &  4.7   &  3.50\\
      1334.53  & 	C~{\sc ii}         &  4.4   &  155.64\\
\hline
\end{tabular}
\end{table}

In order to obtain the  off-limb radiances it is important to accurately determine the position 
of the solar limb. For this purpose we first selected the part of the SUMER spectrum dominated by 
continuum emission around 1230~\AA\ and the emission in the C~{\sc i} line at 1252.21~\AA\ 
(log$T~(K)~=~4.0$). A tilt in the orientation of the grating with respect to the detector causes 
the emission along the SUMER slit to be shifted with the dispersion. This vertical displacement was 
accounted for by using the program delta\_pixel.pro. We cut a further two pixels above 
the identified limb in order to avoid contamination from disk emission. After having subtracted the 
background from our data we fitted each spectral line with a single Gaussian to determine the 
total off-limb flux contribution by that line. The Poisson noise for these radiances were 
also calculated.

\section{DEM analysis}
A DEM distribution gives information about the plasma distribution as a function of temperature along a 
given line-of-sight (LOS). A DEM, $\phi(T)$, is defined as
\begin{equation}
\phi(T) = n_{e}^2~\frac{dh}{dT}
\end{equation}
where $n_{e}$ is the electron density at position $h$ along the LOS at temperature $T$. For an 
optically thin line in ionisation equilibrium, the DEM and the observed intensity ($I_{obs}$) are 
associated to each other by the equation
\begin{equation}
I_{obs} = \int A(x)~G(N,T)~\phi(T)~dT
\end{equation}
where $A(x)$ is the elemental abundance with respect to hydrogen and $G(N, T)$ is the contribution 
function for a given spectral line. This relationship can be 
used to derive the DEM from an observed spectrum \citep{Withbroe1975} using inversion techniques.

\begin{figure}[ht]
\centering
\includegraphics[width=10cm]{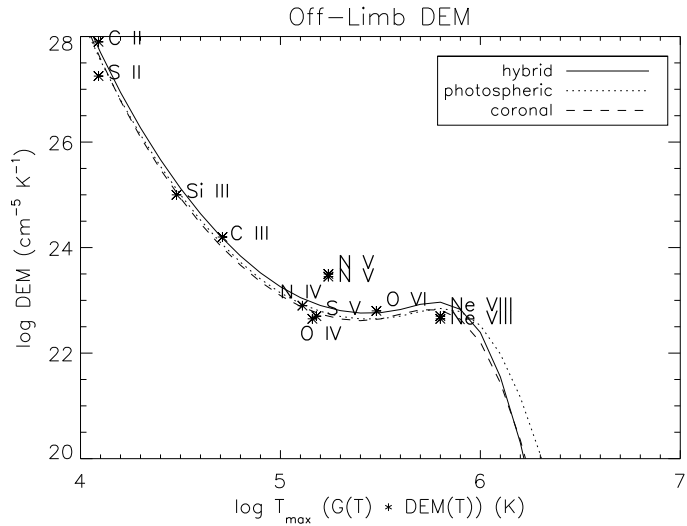}
\caption{The DEM for off-limb data (where spicules are most easily identified) is calculated 
for different abundances using CHIANTI 6.0.1.}
\label{dem}
\end{figure}

A deconvolution method is used to obtain the DEM from measured spectral intensities. In this paper 
we use the CHIANTI atomic database version 6.0.1 \citep{Dere2009}. A procedure available along with 
the CHIANTI package called chianti\_dem.pro is used to derive the DEM distribution. This program 
accepts wavelength, observed intensities and corresponding errors of the spectral lines along with 
pressure/density and elemental abundances corresponding to the observed region of the Sun as inputs. 
The best fit for the DEM can be controlled by changing the ``mesh points'' (an array that specifies the points for the spline that represent the fitted DEM (see CHIANTI: User Guide for further details)).

\section{Results and Discussion}
We have successfully constructed a DEM distribution of spicules for different solar abundances 
that are predefined in the CHIANTI package (see Figure~\ref{dem}). The derived DEM distribution was used to 
analyse the thermal response of the AIA 171~\AA\ channel to spicular emission.

\begin{figure}
\centering
\includegraphics[width=10cm]{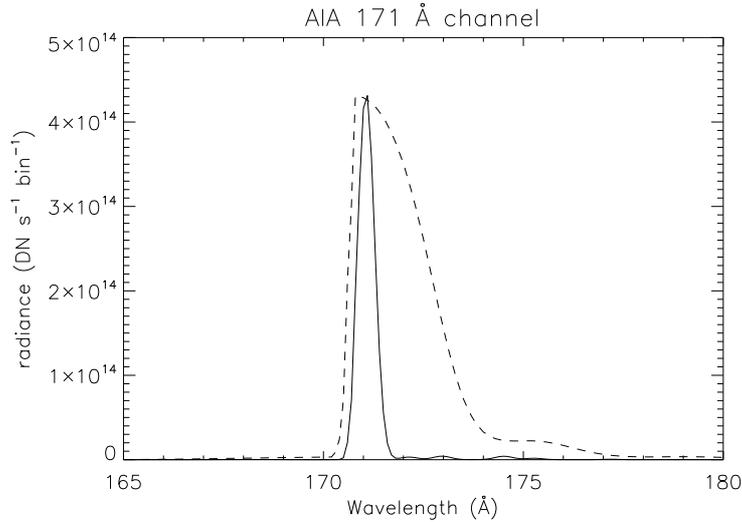}
\caption{The AIA 171~\AA\ passband spectral response from the model DEM (solid line). The wavelength 
response function of this band is plotted with a dashed line.}
\label{171area}
\end{figure}

\begin{figure}
 \centerline{\hspace*{0.015\textwidth}
               \includegraphics[width=0.6\textwidth,clip=]{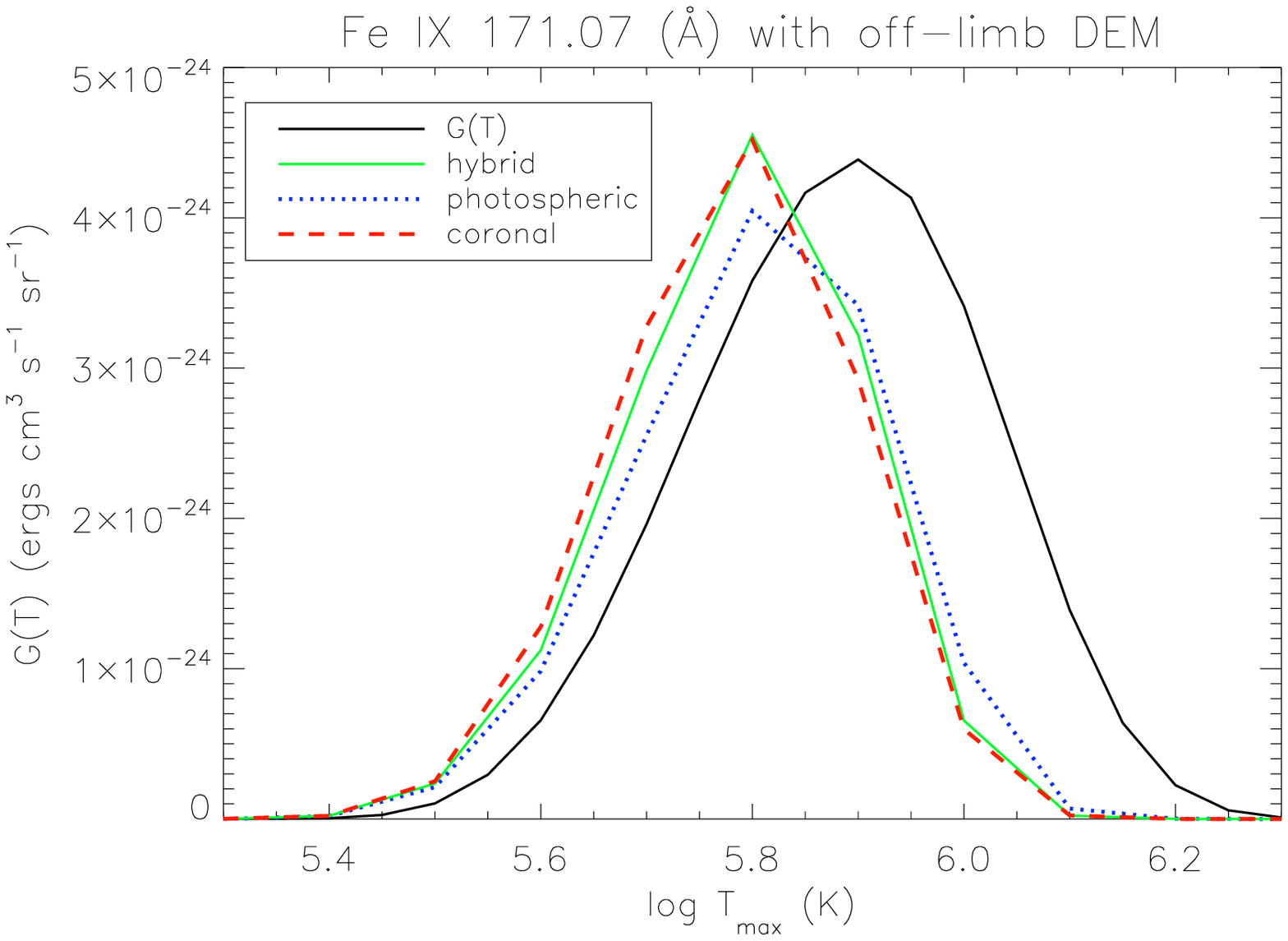}
               \hspace*{-0.03\textwidth}
               \includegraphics[width=0.6\textwidth,clip=]{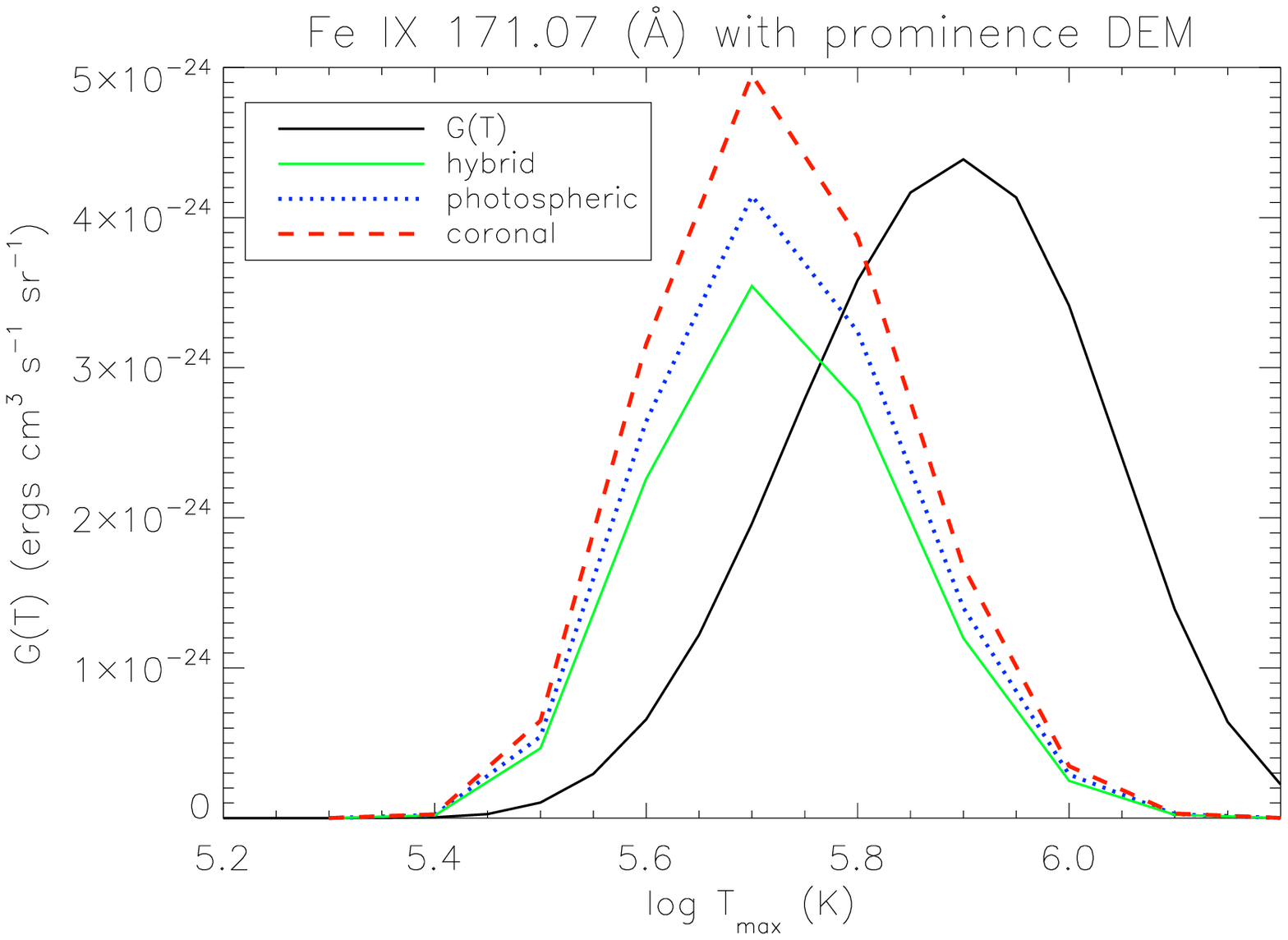}
              }
     \vspace{-0.35\textwidth}   
     \centerline{\bf     
      \hspace{0.40 \textwidth}  \color{black}{(a)}
      \hspace{0.60\textwidth}  \color{black}{(b)}
         \hfill}
     \vspace{0.31\textwidth}    
   \centerline{\hspace*{0.015\textwidth}
               \includegraphics[width=0.6\textwidth,clip=]{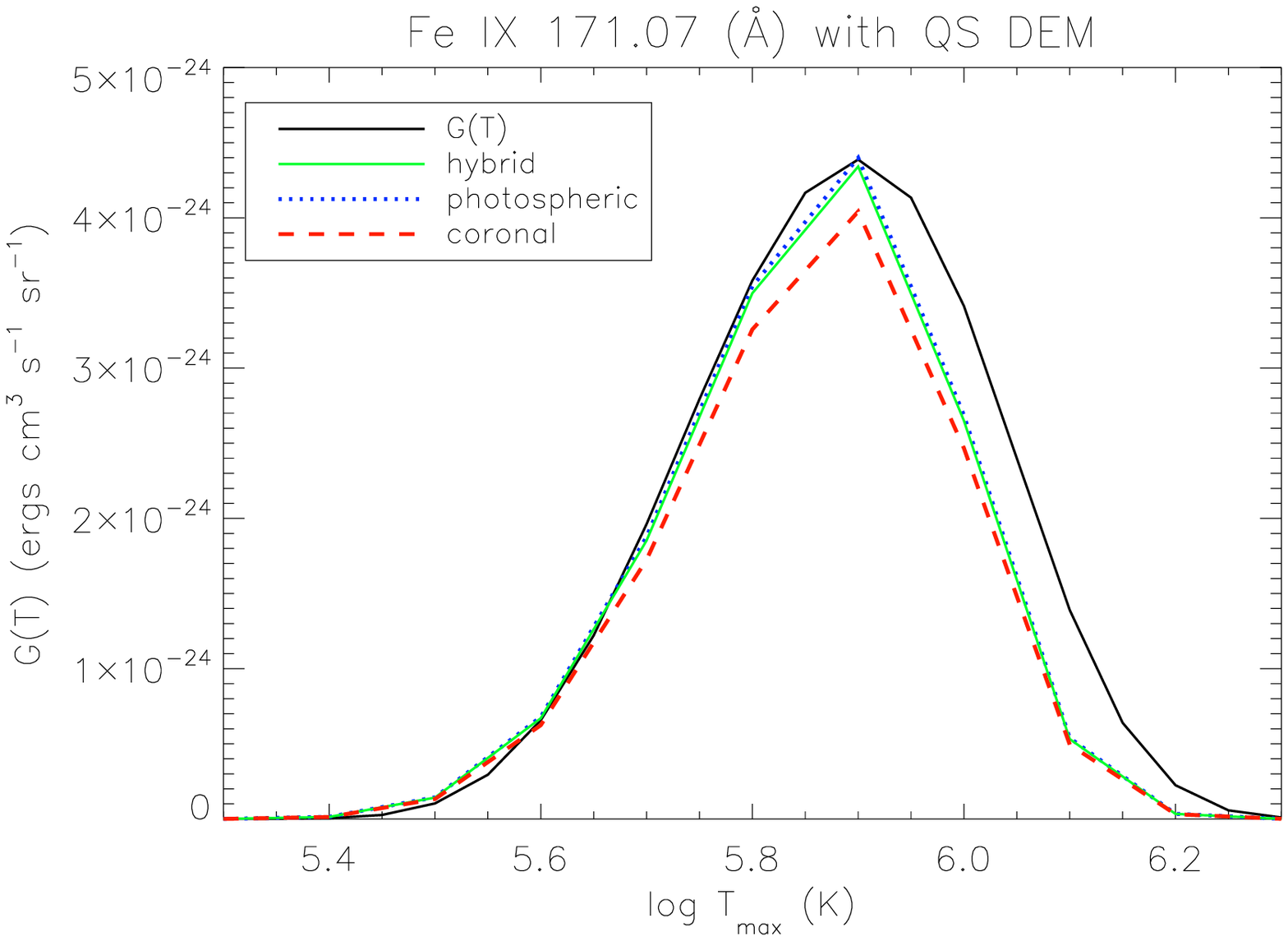}
              }
     \vspace{-0.35\textwidth}   
     \centerline{\bf     
      \hspace{0.7\textwidth} \color{black}{(c)}
         \hfill}
     \vspace{0.31\textwidth}    
              
\caption{The contribution function $G(N,T)$ for Fe~{\sc ix}~171.07~\AA~(black, solid line) is 
plotted along with the normalised product $G(N,T)~\times~DEM(T)$ for three different DEMs (a) off-limb, (b) prominence and (c) quiet Sun and for 
different solar abundances as indicated in the legend.}
   \label{GT171}
   \end{figure}

In  Figure~\ref{171area} we present the spectral response of the AIA 171~{\AA} channel 
from our DEM modeling. We found that the radiance in this channel is dominated by emission from  
the Fe~{\sc ix}~171.07~\AA\ line and has sparingly little contribution from other  lines. However, the product 
of the contribution function and DEM, $G(N,T) \times DEM(T)$ provides a better understanding of 
the formation temperature of this line for a given phenomenon in the solar atmosphere. This 
analysis revealed that, although, the contribution function of the Fe~{\sc ix}~171.07~\AA\ line peaks at 
log$T_{max}~(K)~=~5.9$ (the solid line in Figure~\ref{GT171}(a)), the product $G(N,T) \times DEM(T)$  
has its maximum at log$T_{max}~(K)~=~5.8$ and, therefore, the emission comes from plasma at transition region temperatures. 
 This supports the spectral results of \cite{Madjarska2011}  that  spicules may not attain temperatures greater than 300\,000~K. 

We did a comparative study with prominences since it is known that their plasma properties are similar to that of spicules. As expected
the response of the 171~\AA\ line to the prominence DEM from CHIANTI (see Figure~\ref{GT171}(b)) showed a significant shift in the
emission contribution to log$T_{max}~(K)~=~5.7$ which is even  more pronounced than in the case of spicules. Prominences are
visible in the 171~\AA\ passband of AIA but from Figure~\ref{GT171}(b) we see that the contribution from million degree plasma is very insignificant.


We then made a similar plot using the quiet Sun DEM from CHIANTI (see Figure~\ref{GT171}(c)) and found 
very contrasting results. The peaks of the line contribution function and the product 
$G(N,T) \times DEM(T)$ coincide here, unlike the case with the off-limb DEM and prominence DEM. The differences between 
the three DEMs strongly influence the obtained results. We suggest that a quiet Sun DEM could be affected 
by emission from plasma at widely differing temperatures as it could originate from numerous 
sources like loops, coronal bright points, spicules, filaments etc. in an average quiet Sun region. 

From the graphs in Figure~\ref{GT171} we were able to numerically integrate the values to obtain what percent of plasma seen in the
171~\AA\ filter is above  million degrees. We find that in the case of spicule or prominence observations the contribution is only 3.6\%
and 0.9\%, respectively, whereas for the quiet Sun it is a little over 15\%.

\begin{figure}
\centering
\includegraphics[width=10cm]{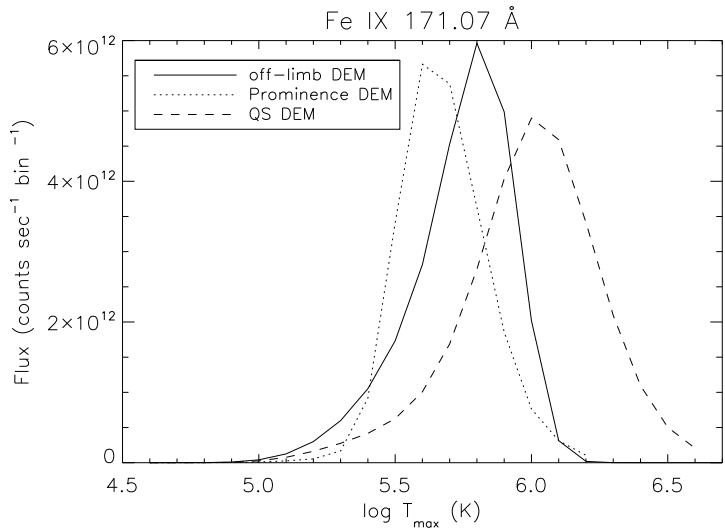}
\caption{A comparison of the Fe~{\sc ix} 171.07~\AA\ line flux as a function of temperature using both off-limb DEM (solid 
line), quiet Sun DEM (dashed line) and prominence DEM (dotted line)}.
\label{emiss171}
\end{figure}

We then obtained the line emission flux as a function of temperature. Using CHIANTI we calculated 
the line emissivity as a function of temperature and then folded it with the QS, prominence and our 
off-limb DEM distributions. Here again we find discrepancies for the three DEMs used as shown in
Figure~\ref{emiss171}. While the emission of Fe~{\sc ix} 171.07~\AA\ in response to the off-limb 
DEM peaks at log$T_{max}~(K)~=~5.8$ (transition region temperature) and that of the prominence DEM is much lower at log$T_{max}~(K)~=~5.6$,
 we find that the maximum flux  using the quiet Sun DEM  occurs at log$T_{max}~(K)~=~6.0$ (coronal temperature). Since these differences exist in 
the use of the Fe~{\sc ix}~171.07~\AA\ line, results pertaining to observations in this line can easily be misrepresented.

The sensitivity of some spectral lines to different DEM has previously been demonstrated by \citet{Brooks2011}. The authors
investigated  spectroheliograms of active region fan loops produced from  Fe~{\sc viii} and Si~{\sc vii} lines. They found striking similarities
in the appearance of the fan loops despite their different formation temperature, log$T~(K)$ = 5.6 and 5.8 respectively. Note that Fe~{\sc viii}
is the main contributor to the AIA~131~\AA\ channel. It has been found that the Fe~{\sc viii}~185.213~\AA\ line is particularly sensitive to the
slope of the DEM, leading to disproportionate changes in its effective formation temperature. This is similar to the behaviour of the Fe~{\sc ix}
line studied here and its impact to the temperature response of the AIA~171~\AA\ channel.

\section{Conclusions}
The aim of the present study was to derive a DEM  distribution for a region dominated by spicular emission, best described by an
off-limb environment above a coronal hole. Until now spicular studies have always been spoken about in context to the quiet Sun.
However, in this paper using spicular, prominence and quiet-Sun DEM distributions, we demonstrated that there is an obvious
difference in how the Fe~{\sc ix}~171.07~\AA\ emission changes in these regions. From the off-limb DEM we find that during spicule
observations the bulk emission in the AIA 171~\AA\ filter is from cool plasma at  log$T~(K)~=~5.8$ to as low as log$T~(K)~=~5.5$ with
only 3.6\% of the plasma being above a million degrees. A similar deduction was made in regard to prominence observations
using this filter where the filter is receptive to emission at log$T~(K)~=~5.7$ to as low as log$T~(K)~=~5.5$ with a meager 0.8\%
being over million degrees. Whereas for a quiet Sun region there is significant emission from plasma at log$T~(K)~=~5.9$ to
log$T~(K)~=~5.55$ with over 15\% being above million degrees. From our study we find that the temperature sensitivity of this filter
depends on the kind of feature that is observed. Although the Fe {\sc ix} line can be formed or has a contribution from plasma
at million degree temperature, our results suggest that spicule observations in this filter cannot be used as a conclusive evidence that these
phenomena are heated to coronal temperatures. We, therefore, emphasise that future studies related to off-limb features, especially
spicules, should adopt the use of the off-limb DEM.  We propose that the quiet Sun can be dominated by plasma from neighbouring
structures (e.g. loops \textit{etc}.) and this could give rise to the differences that we point out here.

Our results are in agreement with the conclusions made by \cite{DelZanna2011} where they use data from active region loops to
investigate spectral line contribution to the EUV AIA channels. The outcome of our work shows that mostly transition region emission
is registered through the AIA 171~\AA\ channel during observations of certain phenomena like spicules and prominences. We need
to further explore the issue raised here and have a better understanding of the thermal characteristics of the observed phenomenon
in order to be able to completely and truthfully exploit the state-of-the-art AIA instrument.

%

%

%

%
 \begin{acks}
We would like to thank the anonymous referee for the important suggestions and comments on this manuscript.
 K.V, M.M, J.G.D thank ISSI for the support of the team ``Small-scale transient phenomena and 
their contribution to coronal heating''. Research at Armagh Observatory is grant-aided by the 
N. Ireland Department of Culture, Arts and Leisure. CHIANTI is a collaborative project involving 
NRL (USA), RAL (UK), and the Universities: College London (UK), of Cambridge (UK), George Mason 
(USA), and of Florence (Italy). The AIA data are courtesy of SDO (NASA) and the AIA consortium. 
The SUMER project is financially supported by DLR, CNES, NASA and the ESA 
PRODEX programme (Swiss contribution). This work was supported via grant 
ST/F001843/1 \& ST/J00135X from the UK Science and Technology Facilities Council.

 \end{acks}

\begin{figure}[ht!]
\centering
\vspace{-1.5cm}
\includegraphics[width=0.99\textwidth]{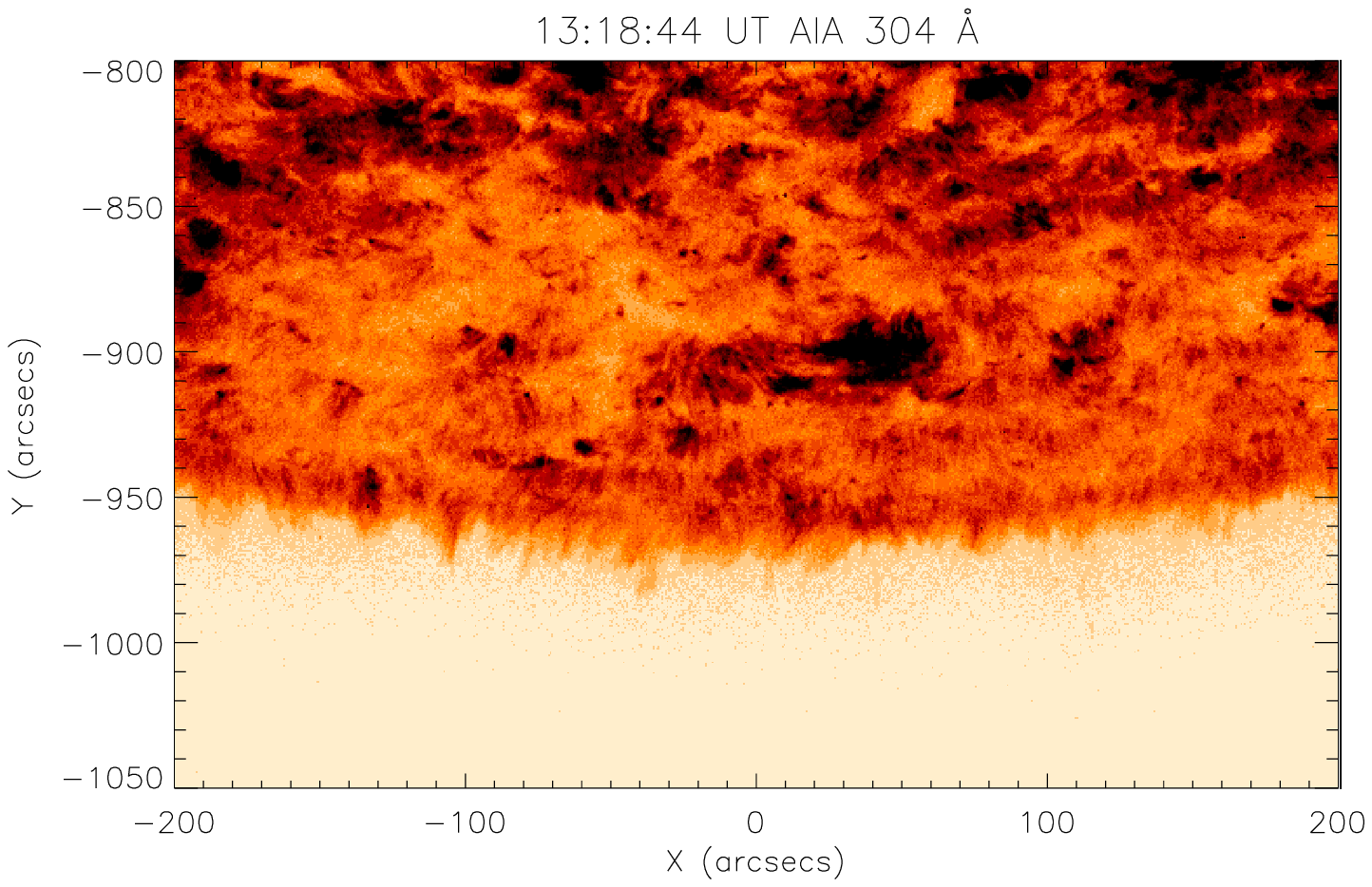}
\vspace{-2cm}

\includegraphics[width=0.99\textwidth]{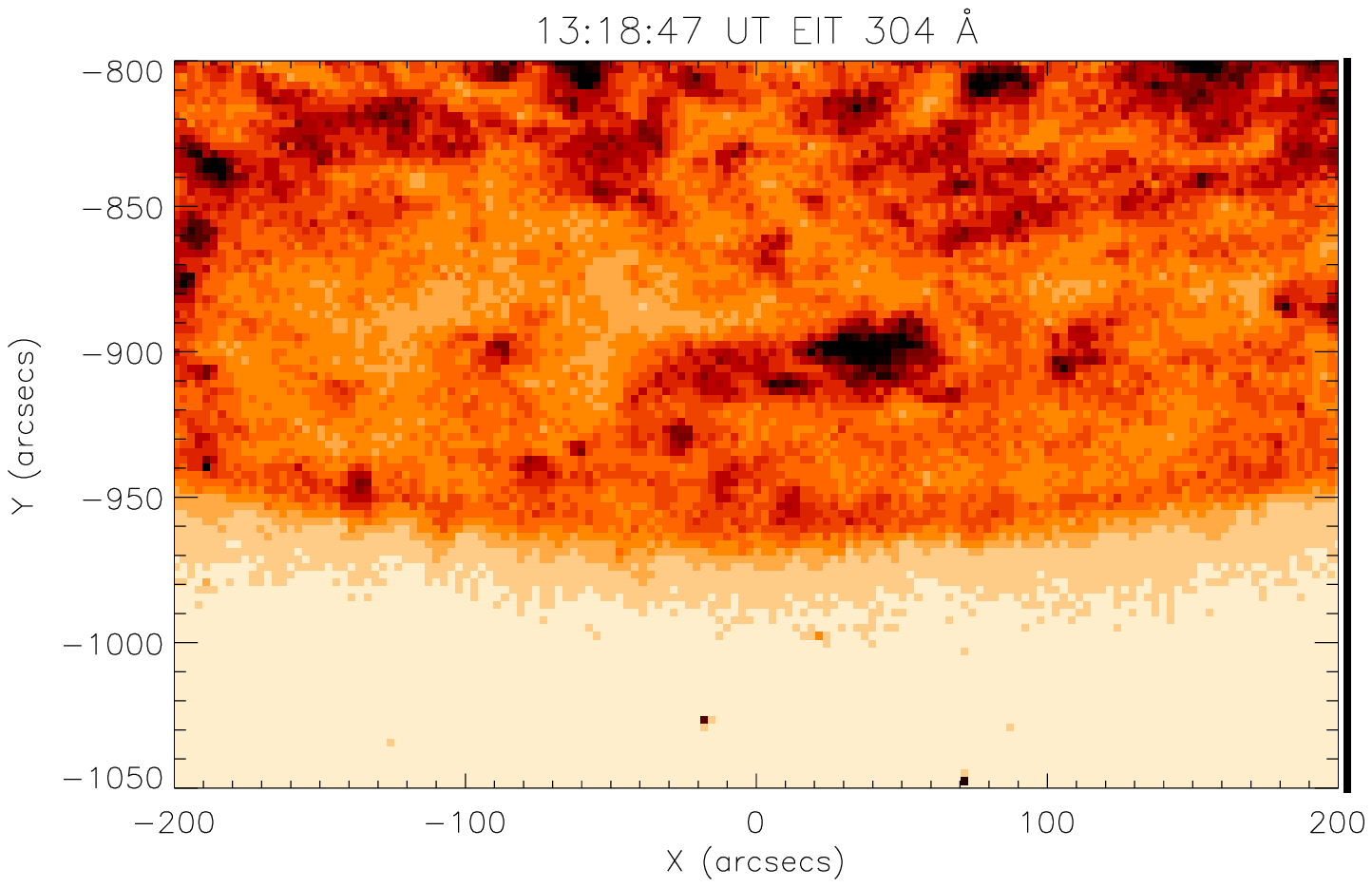}
\vspace{-1cm}
\caption{AIA 304~\AA\ (top) and  EIT 304~\AA\ (bottom) images  (colour table reversed) of a coronal hole region at the South Pole taken on 14 May 2010.}

\label{aia_eit}
\end{figure}

\begin{appendix}
As mentioned in Section~2, coronal hole off-limb data are dominated by spicules. We use 
EIT images (see Figure~\ref{eit_sumer}) to ensure that no coronal bright points are present in 
the SUMER field-of-view since EIT was the only available imager at the time the SUMER reference 
spectra were obtained. However, the EIT images do not distinctly reveal spicular structure due 
to some instrumental limitations: low spatial resolution and high stray-light contribution. In 
order to establish what we see in the EIT image we compare it with co-temporal AIA~304~\AA\ 
image (see Figure~\ref{aia_eit}) which has a better spatial resolution. While the off-limb EIT 
image shows blurred streaks merging together due to its low spatial resolution, the AIA image 
reveals numerous narrow spikes, i.e. spicules, confirming the abundance of spicules in off-limb 
coronal hole data.

\end{appendix}
%
\bibliographystyle{spr-mp-sola-cnd} 
 \bibliography{references}  
%
%
%
%

\end{article} 
\end{document}